\documentclass[%
 reprint,
superscriptaddress,
 amsmath,amssymb,
 aps,
]{revtex4-1}

\usepackage{graphicx}
\usepackage{dcolumn}
\usepackage{bm}

\begin{document}

\newcommand{\mvec}[2]
{
\left(\begin{array}{c}
#1  \\
#2  
\end{array}
\right)
}

\newcommand{\mmat}[4]
{
\left(\begin{array}{cc}
#1  & #2\\
#3  & #4
\end{array}
\right)
}

\newcommand{\mvecthree}[3]
{
\left[
\begin{array}{c}
#1  \\
#2  \\
#3  
\end{array}
\right]
}
\newcommand{\defto}{\stackrel{\rightharpoonup}{=}}
\newcommand{\deffrom}{\stackrel{\leftharpoonup}{=}}
\newcommand{\mi}{{\mathrm i}}
\newcommand{\cmt}[2]{{[}#1,#2{]}}
\newcommand{\acmt}[2]{{\{}#1,#2{\}}}


\title{Bulk-edge correspondence with generalized chiral symmetry}

\author{Tohru Kawarabayashi}
\affiliation{Department of Physics, Toho University,
Funabashi, 274-8510 Japan}

\author{Yasuhiro Hatsugai}
\affiliation{Department of Physics, University of Tsukuba, Tsukuba, 305-8571 Japan}

\date{\today}

\begin{abstract}
The bulk-edge correspondence in topological phases is extended to systems with the generalized chiral symmetry, 
where the conventional chiral symmetry is broken. In such systems, we find that 
the edge state exhibits an unconventional behavior in the presence of the symmetry breaking by the mass, which is explored explicitly 
in the case of a deformed Su-Schrieffer-Heeger model.   
The localization length of the edge states diverges at a certain critical mass, where the edge 
state touches the bulk band. 
The edge state is specified by an imaginary wave vector that becomes real at the touching energy.
\end{abstract}


\maketitle

\section{Introduction}

Since the discovery of the topological insulators and superconductors \cite{HK,QZ}, topological states of 
matter has been one of the central issues of the condensed matter physics. 
When the bulk topological invariant has a non-trivial value, the topological edge states emerge at the edges or boundaries of the system. This relationship between the bulk states and the edge states has been the hallmark of the topological phases of matter, which is 
called the bulk-edge correspondence originally introduced for the quantum Hall
effect \cite{Hatsugai}. 
In classifying the topological phases, the chiral symmetry has been one of the important symmetries \cite{AZ,RSFL,HA1,CTSR}.
In particular, it has been shown that the chiral symmetry protects the zero-energy edge states at the boundary of the topological systems \cite{RH}. The concept of the bulk-edge correspondence in chiral symmetric systems has recently been extended to non-hermitian systems \cite{YW,YM}.

Here in the present paper, we show that the bulk-edge correspondence can be extended to systems respecting
an extension of the chiral symmetry, which we call the generalized chiral symmetry \cite{KHMA,HKA,KAH}. This can be achieved by 
the algebraic deformation of the chiral symmetric Hamiltonian in which the generalized chiral symmetry, introduced originally for the 
characterization of the tilted Dirac fermions in two dimensions \cite{TSTNK,KKS,KKSF,KSFG,KNTSK,GFMP,MHT}, is always preserved. 
The deformation can be performed exactly in lattice models  as well as in continuum models.
The key ingredient is that the number of zero modes is an invariant of the deformation and this fact has enabled us to extend  
the topological protection of the doubling of the massless Dirac fermions on two-dimensional lattice models, 
which is understood as a consequence of the chiral symmetry \cite{NN,HFA,Hatsugai2},  to the tilted Dirac fermions on a lattice model respecting the generalized chiral symmetry \cite{KAH2}.  In the deformed systems, we find that 
the edge state shows an instability against the symmetry breaking by the mass. This is in sharp contrast to the case with the conventional chiral symmetry where the edge state is always robust against the symmetry breaking by the mass. This unconventional behavior of 
the edge state is explored in detail for the one-dimensional Su-Schrieffer-Heeger model.

The one-dimensional Su-Schrieffer-Heeger (SSH) model \cite{SSH}, introduced for analyzing the soliton state of polyacetylene, is a simple tight-binding model with a bond alternation induced by the electron-lattice coupling.  Fundamental physical phenomena such as topological excitations in one dimension as well as the charge fractionalization associated with them \cite{JR,SS,GW,RM,JS,HKSS}  have 
been investigated based on the SSH model. It has served further as a prototypical model for illustrating the bulk-edge correspondence in the presence of the chiral symmetry \cite{RH,YW,YM}. In the recent progress of the experimental technique, 
the SSH model itself has been realized  in an atom-optical system as well as in engineered atomic chains
where the topological state  has been experimentally confirmed \cite{MAG,DOHL,HKODL}.

The paper is organized as follows. We introduce,  in section II, the general theoretical frame work for the generalized chiral symmetry  and 
the effect of the symmetry breaking by the mass. In section III, we explore the unusual bulk-edge correspondence 
with the generalized chiral symmetry for  deformed Su-Schrieffer-Heeger models. Section IV is devoted to the summary.

\section{General Formalism}

\subsection{Generalized Chiral Symmetry}
To discuss the bulk-edge correspondence in the topological phases, we consider systems with edges or boundaries. The algebraic deformation
for generating a series of systems with the generalized chiral symmetry can be  generally applicable to such systems with edges or boundaries \cite{KAH}. 
The generalized chiral symmetry exists when the Hamiltonian $H$ satisfies 
$$
 \gamma^\dagger H \gamma = -H,
$$ 
with $\gamma^2 =1$,
where the generalized chiral operator $\gamma$ is not necessarily hermitian \cite{KHMA,HKA}. The generalized chiral symmetry 
is an extension of the conventional chiral symmetry since it reduces to the conventional one when the operator $\gamma$ is hermitian.
The generalized chiral symmetry can be defined exactly for lattice Hamiltonians 
as well as for effective low-energy Hamiltonians \cite{KAH}.

To be specific, we consider the case where the original lattice Hamiltonian is bipartite.
A bipartite Hamiltonian can generally be expressed  as 
$$
 H_{\rm c} = \mmat{O}{D}{D^\dagger}{O},
$$ 
in a basis $(\Psi_{a_1},\ldots ,\Psi_{a_N},\Psi_{b_1},\ldots , \Psi_{b_N})$ where $\Psi_{a_n(b_n)}$ denotes the basis of  the A(B) sublattice in the $n$-th unit cell. 
Here $D$ is a $N$ by $N$ matrix and $N$ denotes the number of the unit cells in the system.
In this case, the conventional chiral operator is given by  
$$
 \Gamma =\mmat{I_N}{O}{O}{-I_N} = \sigma_z \otimes I_N  ,
$$
since it satisfies the relation $\Gamma H_{\rm c}\Gamma = -H_{\rm c}$ and $\Gamma^2=1$. Here $I_N$ stands for the $N$ by $N$ 
identity matrix.
The series of lattice Hamiltonians $H_\tau(q)$ respecting the generalized chiral symmetry can be generated from the chiral symmetric lattice Hamiltonian 
$H_{\rm c}$ 
by the algebraic transformation as 
\begin{equation}
 H_\tau(q) = T_\tau(q)^{-1} \ H_{\rm c} \ T_\tau(q)^{-1}.
 \label{deformation} 
\end{equation}
with  
$$
 T_\tau(q) = \exp(q\bm{\tau}\cdot\bm{\sigma}/2) \otimes I_N  
$$
where $\bm{\tau} = (\tau_x,\tau_y,\tau_z)$ is a three-dimensional real and unit vector  and 
$\bm{\sigma}=(\sigma_x,\sigma_y, \sigma_z)$ are Pauli matrices. The parameter $q$ is assumed to be real and 
thus $T_\tau(q)$ is an hermitian matrix with $\det T_\tau(q)=1$.
We then define the generalized chiral operator $\gamma$ as 
$$
 \gamma = T_\tau(q) \ \Gamma\  T_\tau(q)^{-1}.
$$
It is straightforward to see that $\gamma^2=1$ and 
$\gamma^\dagger H_\tau(q) \gamma = -H_\tau(q)$. The Hamiltonians deformed by the transformation (\ref{deformation}) therefore always 
respect the generalized chiral symmetry. This hyperbolic transformation has the same form as the Lorentz boost (Appendix A).

Inversely, if we require the generalized chiral symmetry for a lattice model with the bipartite structure, it has been shown
that the lattice model can be transformed back to a chiral symmetric lattice model \cite{KAH2}. For the case of 2 by 2 
matrices, a matrix $\gamma_2$ satisfying $\gamma_2^2=1$ can be expressed in the form
$$
 \gamma_2 = \exp(q\bm{n}_1\cdot\bm{\sigma}/2) (\bm{n}_0\cdot \bm{\sigma})  \exp(-q\bm{n}_1\cdot\bm{\sigma}/2),
$$
where $\bm{n}_0$ and $\bm{n}_1$ are real vectors with $\bm{n}_0^2=\bm{n}_1^2=1$ and $\bm{n}_0\cdot\bm{n}_1=0$. 
The generalized chiral operator $\gamma_s$ can thus be expressed generally in the real space as
$$
 \gamma_s =\gamma_2 \otimes  I_N  = S_{n_1}(q)\  \Gamma' \ S_{n_1}(q)^{-1}
$$
with 
$$
 S_{n_1}(q) =    \exp(q\bm{n}_1\cdot\bm{\sigma}/2)\otimes I_N , \quad \Gamma' = (\bm{n}_0\cdot \bm{\sigma})  \otimes I_N.
$$
When the Hamiltonian $H$ respects the generalized chiral symmetry as
$\gamma_s^\dagger H \gamma_s = -H$,
we can then define an inverse transformation as 
\begin{equation}
 H_{\rm c}' = S_n(q) H S_n(q).
 \label{inverse_deformation}
\end{equation}
It is then verified that 
$H'_{\rm c}$ is indeed chiral symmetric,  because it satisfies the relation $\Gamma' H_{\rm c}' \Gamma' = -H'_{\rm c}$ with
$\Gamma' = (\bm{n}_0\cdot \bm{\sigma})$ and $(\Gamma')^2 = 1$.

It should be noted that the number of the zero energy states is an
invariant of the transformation and its inverse. If we have a zero energy state $\psi_0$, for instance, of the original Hamiltonian $H_{\rm c}$,
namely $H_{\rm c}\psi_0 = 0$, then it is easy to see that the state defined by $T_\tau(q) \psi_0$ is also a zero energy state of 
$H_\tau(q)$ since $H_\tau(q) (T_\tau(q)\psi_0) = T_\tau(q)^{-1} H_{\rm c} \psi_0 =0$. Taking into account that $\det T_\tau(q)=1$, we can safely conclude 
that the number of zero energy states is an invariant of the deformation.  

If the original system $H_{\rm c}$ is  topologically non-trivial, we have the topological edge states 
at the open boundary of the system, which are the zero energy states because of the chiral symmetry. 
It is to be remarked, however, that the energy of the edge state localized at the boundary 
becomes exactly zero only in the thermodynamic limit ($N \to \infty$) where 
the mixing between edge states at both boundaries becomes negligible.
In a conventional approach, therefore, a semi-infinite system with one
boundary has been considered to define the edge states as 
the zero energy states.  
Here in the present paper, we adopt an alternative approach to define the edge states based on the exact zero energy states for a finite system ($N<\infty$). 

Instead of a semi-infinite system, we consider simply a finite system ($N<0$) having the left and the right boundaries at both ends of the system. 
To define the edge state, for example, at the left boundary, we first 
modify the Hamiltonian at the right end of the system so that there exist exact zero energy states even for a finite $N$. 
This can be achieved rather easily as we demonstrate for the SSH model in section III. We then take the 
thermodynamic limit $N\to \infty$ where one of the zero modes becomes the edge state at the left boundary. It is to be noted that, in the thermodynamic limit, 
the local modification of the Hamiltonian in the vicinity of one end of the system should be 
negligible for the edge state localized at the other end. Though this approach gives the same result as the conventional one, 
it has an advantage that we can safely assume the existence of the exact zero energy states
even in the present real-space formalism where the system is described by a large but finite-size matrix.
   
In numerical analyses for finite systems, the deviation of the energy of the edge state from zero should be exponentially small and thus 
can be negligible
when the system-size ($N$)  is much larger than the localization length of the edge state.

Since the deformation  preserves the number of zero modes, 
the number of the edge states is also an invariant of the deformation. This clearly  
suggests that the bulk topological invariant should also be preserved within the present deformation. 
This property of the deformation also leads to the fact that if the original chiral symmetric model has an energy gap in the 
bulk spectrum, it remains open and 
never closes in any deformed models with arbitrary $q$. 
Note that because of the chiral symmetry, if the gap exists, it should be open symmetrically around zero 
energy ($E=0$) in the original model. In the deformed Hamiltonian generated by the transformation (\ref{deformation}), 
the zero energy states 
should never appear in the bulk spectrum and the positive/negative energy states should remain positive/negative for any $q$.

\subsection{Symmetry Breaking}

Let us discuss the effect of the symmetry breaking by the mass term 
$$
 H_m = m\Gamma.
$$
To be specific, we assume $m$ is positive.
We find that the robustness of the edge states against the mass term for $q\neq 0$ turns out to be 
quite different from that for $q=0$. To see this, it is instructive to revisit the robustness of the edge state 
for $q=0$. In this case, the Hamiltonian is chiral symmetric for $m=0$, and therefore the zero energy edges states
can be  expressed as an eigenstate of the chiral operator. This leads to the fact that the
energies of the edge states are exactly given by $E_{\rm edge} (m) = \pm m$ and the corresponding eigenstates 
are independent of $m$. The edge state localized at the boundary therefore never disappears for any value of $m$.

For deformed systems ($q\neq 0$), the behavior of the edge states in the presence of the mass term is qualitatively different. 
Let us recall the general theoretical framework \cite{KAH} for the energy eigenvalues 
of deformed systems with the generalized chiral symmetry. To discuss systems with open boundaries, 
we proceed  the real space representation in a 
basis $(\Psi_{a_1},\ldots,\Psi_{a_N},\Psi_{b_1},\ldots,\Psi_{b_N})$. 
The deformed Hamiltonian in the presence of the mass term is then defined by
$$
 H_\tau^{(m)}(q) = T_\tau(q)^{-1} (H_{\rm c}+m\Gamma) \   T_\tau(q)^{-1} .
$$
Since the operator $T_\tau(q)$ with $\bm{\tau} = (0,0,1)$ induces no changes for $H_{\rm c}$, we confine 
ourselves to the case $\bm{\tau}=(\cos\theta, \sin\theta,0)$. We then have
$$
  H_\tau^{(m)}(q)= H_\tau(q) +m\Gamma
$$
since we have a relation $\Gamma\  T_\tau(q) = T_\tau(-q) \Gamma$. Eigenvalue equations are given by
$$
 H_\tau^{(m)}(q) \psi_E^{(m)} = E \psi_E^{(m)},
$$  
where $E$ depends on $m$. Multiplying $H_\tau^{(m)}(-q)=H_\tau^{(m)}(q)+I_2 \otimes (e^{i\theta}D+e^{-i\theta}D^\dagger)\sinh q $, we have \cite{KAH}
\begin{eqnarray*}
 \lefteqn{H_\tau^{(m)}(-q)H_\tau^{(m)}(q) \psi_E^{(m)} }\\
 & =&T_\tau(q)  (H_{\rm c}^2 +m^2)T_\tau(q)^{-1}  \psi_E^{(m)}\\
 &=& E(E+ I_2 \otimes (e^{i\theta}D+e^{-i\theta}D^\dagger)\sinh q )\psi_E^{(m)},
\end{eqnarray*}
which leads to
$$
 (H_{\rm c}^2 - I_2 \otimes E (e^{i\theta}D+e^{-i\theta}D^\dagger)\sinh q  +m^2)  \Phi_E^{(m)} = E^2\Phi_E^{(m)}
$$
with $\Phi_E^{(m)} = T_\tau(q)^{-1}  \psi_E^{(m)}$. Completing the square, we have 
$$
 [(H_{\rm c}(q,E))^2 + m^2_R] \Phi_E^{(m)} = E^2 \Phi_E^{(m)}
$$
with
\begin{eqnarray*}
 H_{\rm c}(q,E) &=& \mmat{O}{D(q,E)}{D^\dagger(q,E)}{O}, \\
 D(q,E) &\equiv & \frac{1}{\cosh q}(D-e^{-i\theta}E\sinh q)
\end{eqnarray*}
and $m_R \equiv m/\cosh q$.
We have therefore generally $E^2 \geq m_R^2$. 
The condition under which the eigenstate with an eigenvalue $E = m_R$ in the form 
\begin{equation}
 \psi_{E=m_R}^{(m)} = T_\tau(q) \mvec{\phi_+^m}{0}
 \label{form1}
\end{equation}
exists, is given by \cite{KAH} 
\begin{equation}
 (D^\dagger -m_R e^{i\theta} \sinh q) \phi_+^m =0 .
 \label{leftedge}
\end{equation}

For the eigenstate $\psi_{E=-m_R}$ with the energy $E=-m_R$ in the form
\begin{equation}
 \psi_{E=-m_R}^{(m)} = T_\tau(q) \mvec{0}{\phi_-^m},
 \label{form2}
\end{equation} 
the  condition is similarly given by 
\begin{equation}
 (D +m_Re^{-i\theta}\sinh q)\phi_-^m =0.
 \label{rightedge}
\end{equation}
Note that the eigenstate $\psi_{E=\pm m_R}^{(m)}$, if it exists, is also the eigenstate of the generalized chiral operator $\gamma$
with the eigenvalue $\pm 1$ and reduces to the eigenstate of the conventional chiral operator $\Gamma$
in the limit as $q \to 0$.

The state $\psi_{E=\pm m_R}^{(m)}$ with the energy $\pm m_R$ is apparently a candidate for the edge state of the topological system 
because it is connected to the zero mode in the bulk gap in the limit as $m \to 0$. We therefore look for an eigenstate with $E=\pm m_R$ 
with the form $\psi_{E=\pm m_R}^{(m)}$ decaying exponentially from one end of the system to the other. To discuss such a 
state localized  at one end of the system, it is  allowed to modify the original Hamiltonian locally at the other end of the system, since 
the amplitude of the state is exponentially small there and the effect of the modification should be negligible in the
thermodynamic limit. The modification to obtain, for instance, an exact eigenstate $\psi_{E=m_R}^{(m)}$ is performed 
so that the above condition (\ref{leftedge}) is satisfied by a non-zero solution $\phi_+^m$ even for a finite system.
This can be achieved, practically, by a modification that reduces 
the rank of the matrix $(D^\dagger -m_Re^{i\theta}\sinh q)$ from $N$ to $N-1$.
Though such a modification certainly depends on the details of the Hamiltonian, it is realized generally as long as it is composed of local operators. 
With this modification, we can construct an exact eigenstate $\psi_{E=\pm m_R}^{(m)}$ decaying 
exponentially from one end to the other even for a finite $N$, with which we define the edge state in the limit as $N\to \infty$.

Exponentially decaying states can be understood as  plane waves with complex wave numbers. 
If we  adopt the expression $\phi_\pm^m[j] = e^{ikr_j}u_{\pm, j}^m(k)$ with $u_{\pm,j}^m(k) = u_{\pm,j+\alpha}^m(k)$ where $r_j$ denotes the spatial coordinate 
of the site $j$, which is assumed to increase from left to right, and $\alpha$ is the number of the A(B) site in the unit cell.
The condition (\ref{leftedge})  
for the state $\psi_{E=m_R}^{(m)}$ then can be expressed, as
$$
 (d^\dagger(k) - m_R e^{i\theta}\sinh q )u_+^m(k) =0,  
$$
where $d(k)$ is a complex matrix in the Hamiltonian expressed in the momentum space as
$$
 H_{\rm c}(k) = \mmat{0}{d(k)}{d^\dagger(k)}{0}, 
$$
and $u_+^m(k) = {}^t(u_{+,1}^m(k), \ldots, u_{+,\alpha}^m(k))$.
Note that $d(k)$ is a matrix for a multi-band system ($\alpha >1$). The wave number $k$ is therefore determined by 
$$
 \det[d(k)^\dagger- m_Re^{-i\theta}\sinh q]=0.
$$
Similarly,  for the state $\psi_{E= -m_R}^{(m)}$, the condition (\ref{rightedge}) becomes
$ (d(k) + m_R e^{-i\theta}\sinh q)u_-^m(k) =0$. 
The momentum $k$ is then determined by $\det[d(k) + m_R e^{-i\theta}\sinh q]=0$.

When the above equations have a solution with a wave number $k$ with a non-zero imaginary part, 
${\rm Im}(k) \neq 0$,
the solution is an exponentially decaying (diverging) state which corresponds to the edge state localized at the left(right) boundary of 
a semi-infinite system.
The imaginary part of $k$ must be positive(negative) for an edge state localized at the left(right) boundary, so that 
the edge states are normalizable.
In general, the behavior of the imaginary part of $k$ as a function of the mass $m$ determines the instability of the edge state.  
If  ${\rm Im}(k)$ becomes zero at a certain value $m_c$ of the mass, the edge state at $m=m_c$ is no longer localized  at one boundary of the system and 
is expected to become a bulk state.

As we shall see in the following, for the case of the one-dimensional SSH model with $\alpha=1$, these eigenstates $\psi_{E=\pm m_R}$ are indeed the 
edge states exponentially localized at the boundaries of the system in the presence of the mass term, 
where the sign of the imaginary part of the wave number determines the position of the edge state (left or right boundaries).
It is clearly demonstrated that they merge into the bulk band and disappear
at a certain critical mass where the imaginary part of the wave vector $k$  becomes exactly zero.

\section{Deformed SSH Model}

\subsection{Topological Invariant and Edge states}

\begin{figure}[h]
\includegraphics[scale=0.7]{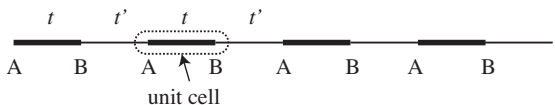}
\caption{
The SSH  model in real space. The transfer integral in the unit cell is denoted by $t$ and that between the 
unit cells is denoted by $t'$. 
\label{ssh_o}
}
\end{figure}

Here we consider deformations of the one-dimensional Su-Schrieffer-Heeger (SSH) model \cite{SSH}, 
which is described by
the Hamiltonian
$$
 H_{\rm SSH} =\sum_{n} t a_n^\dagger b_n + t' a_{n+1}^\dagger b_n + {\rm h. c.},
$$
where $a_n(b_n)$ denotes the annihilation operator of an electron on the sub-lattice  A(B)  
in the $n$-th unit cell (Fig. \ref{ssh_o}). This Hamiltonian respects the chiral symmetry and can be expressed as 
$$
 H_{\rm SSH} = \mmat{0}{D}{D^\dagger}{0}
$$
in the basis $(\Psi_{a_1},\ldots ,\Psi_{a_N},\Psi_{b_1},\ldots , \Psi_{b_N})$ where the non-zero matrix elements of the 
off-diagonal matrix $D$ are given by 
$D_{i,i} = t$ for $i=1,\ldots, N$ and $ D_{i+1,i} = t'$ for $i=1,\ldots, N-1$. The symmetry breaking mass term is defined by
$$
 H_m = m\Gamma = m\sum_{n=1}^N (a_n^\dagger a_n - b_n^\dagger b_n).
$$
In the momentum space, the bulk Hamiltonian with the mass term can be expressed as 
$$
 H_{\rm SSH}(k) + m\sigma_z 
 = \mmat{m}{d(k)}{d^*(k)}{-m},
$$
with $d(k) = t+t'e^{-ik}$. Here we adopt the lattice constant as a unit of length. 
The bulk energy dispersion is therefore given by 
$$
 E (k) = \pm\sqrt{|d(k)|^2+m^2} 
$$
For $m=0$, this model becomes topologically non-trivial when $t' > t$ exhibiting topological edge states at $E=0$ if the 
system has open boundaries,  while it is trivial when $t'<t$ having no edge state at the open boundary \cite{RH,CTSR}. 
In the presence of the mass, the wave function of the edge states remains the same as that for $m=0$ while 
their the energies becomes $\pm m$. 
The edge state with the conventional chiral symmetry therefore never goes into the bulk band since $E(k)^2 > m^2$ as long as $|d(k)|>0$,
nor disappears at a certain finite value of the mass because the wave function itself is independent of $m$.

The deformed SSH model  is defined by the algebraic transformation
$$
 H_\tau (q) =  T_\tau(q)^{-1}\  H_{\rm SSH}\  T_\tau(q)^{-1}
$$ 
with $\bm{\tau}= (\cos \theta , \sin \theta,0)$, which can be represented generally as \cite{KAH}
$$
 H_\tau(q) = -\frac{\sinh q}{2}(e^{i\theta}D+e^{-i\theta}D^\dagger) \otimes I_2 + \mmat{O}{D_q}{D_q^\dagger}{O}
$$
with 
$$
 D_q \equiv D\cosh^2\frac{q}{2} +e^{-2i\theta}D^\dagger \sinh ^2 \frac{q}{2}.
$$
The present deformation of the Hamiltonian in real space can be applied to systems with boundaries.
The deformation preserves the number of zero energy states and therefore preserves the number of the 
edge states of the topological phase in the original SSH model. The deformed system is thus topologically 
non-trivial and has edge states for the case of $t'>t$.

This can be confirmed by evaluating the bulk topological invariant in the system with the translational invariance. 
In such a system, the deformed Hamiltonian can be  given in the momentum space as
$$
 H_\tau(q) = -\sinh q \ {\rm Re}[e^{i\theta}d(k)] I_2 + \mmat{0}{d_{q,\theta}(k)}{d_{q,\theta}^*(k)}{0}
$$ 
with
$d_{q,\theta}(k) = e^{-i\theta}\{{\rm Re}[e^{i\theta}d(k)] \cosh q +i {\rm Im}[e^{i\theta}d(k)]\}$ where ${\rm Re(Im)}[z]$ stands for
the real(imaginary) part of a complex number $z$.
With this expression, it is straightforward to verify that the winding number, which is the topological invariant of the topological phase
of the SSH model \cite{RSFL,RH}, is indeed an invariant of the deformation (see Appendix B).  In real space, the deformed system
can be realized as a ladder system with the next-nearest neighbor transfer-integrals which exhibits various types of 
band structure including a flat band as well as an indirect band gap (Appendix C).

Now let us discuss the symmetry breaking in the deformed systems. 
The deformed SSH model with the mass term is given by
$$
 H_\tau^{(m)}(q) \equiv  T_\tau(q)^{-1}(H_{\rm SSH} + m\Gamma) T_\tau(q)^{-1}
 = H_\tau(q) + m\Gamma.
$$
To discuss the edge states, we consider a finite system with $N$ unit cells having an A site at its left boundary (Fig. \ref{ssh_o}) and examine whether the state with 
the form $\psi_{E=m_R}^{(m)}$  can describe the edge state at the left boundary decaying exponentially toward the right (${\rm Im}(k) >0$).
For the SSH model, the condition (\ref{leftedge}) can be written explicitly as,   
$$
 (t-m_Re^{i\theta}\sinh q)\phi_+^m[n]+t'\phi_+^m[n+1] =0, 
$$
for $n=1,2, \ldots , N-1$, with 
\begin{equation}
 (t-m_Re^{i\theta}\sinh q)\phi_+^m[N]=0
 \label{bc}
\end{equation}
at the right boundary. Here the element of the $n$-th unit cell is denoted by $\phi_+^m[n]$ so that $\phi_+^m = 
{}^t(\phi_+^m[1],\phi_+^m[2], \ldots, \phi_+^m[N])$.  The index $n$ for the $n$-th unit cell is assumed to increase from left to right where  
the unit cell at the left boundary is denoted by $n=1$. 
For a finite $N$, no physical solution is allowed due to the condition (\ref{bc}), suggesting that the 
edge state having exactly the energy $E=m_R$  does not exist in a finite system.

We then perform a modification of $H_{\rm SSH}$ at the right boundary of the system to define a modified Hamiltonian $\tilde{H}_{\rm SSH}^+$ as 
$$
 \tilde{H}_{\rm SSH}^+ = H_{\rm SSH} +\{(m_Re^{-i\theta}\sinh q - t )a_N^\dagger b_N + {\rm h.c.}\},
$$
so that the coefficient of $\phi_+^m[N]$ in (\ref{bc}) vanishes. This modification is nothing but the replacement of the local Hamiltonian $\{t a_N^\dagger b_N + {\rm h.c.}\}$ with $\{(m_Re^{-i\theta}\sinh q) a_N^\dagger b_N + {\rm h.c.}\}$  at the $N$-th unit cell. 
With this modification at the right boundary, the $(N,N)$ element of the matrix $\tilde{D}$ defined as 
$$
 \tilde{H}_{\rm SSH}^+ = \mmat{O}{\tilde{D}}{\tilde{D}^\dagger}{O}
$$ 
becomes  $\tilde{D}_{N,N} = (m_R e^{-i\theta} \sinh q)$, while other elements of $\tilde{D}$ are the same as
$D$. 
Note that the modification affects only on the $(N,N)$ element of the matrix $D$.
It is then easy to verify that the rank of the matrix $[\tilde{D}^\dagger - m_R e^{i\theta} \sinh q ]$ is reduced to  $N-1$ and hence the non-zero 
solution with the energy $m_R$ can exist even for a finite $N$. More explicitly, the equations become
$$
 (t-m_Re^{i\theta}\sinh q)\tilde{\phi}_+^m[n]+t'\tilde{\phi}_+^m[n+1] =0, 
$$ 
for $n=1, \ldots, N-1$, with which the eigenstate $\tilde{\psi}_{E=m_R}^{(m)}$ with the energy $E=m_R$ for $\tilde{H}_{\rm SSH}^+$ is 
given by 
$$
 \tilde{\psi}_{E=m_R}^{(m)} = T(q) \mvec{\tilde{\phi}_+^m}{0}.
$$
Note that the equations for $\tilde{\phi}_+^m$ are exactly the same as those for $\phi_+^m$ except the condition (\ref{bc}).

For the modified system, we can construct  
a non-zero solution decaying(diverging) exponentially from the left boundary to the bulk even for a finite system as 
$\tilde{\phi}_+^m[n] \propto (-1)^{n-1} r_+^{n-1}$ with    
$$
 r_+ \equiv  \frac{t-m_Re^{i\theta}\sinh q }{t'}.
$$
Note that the energy of the state is exactly given  by $m_R$ even for a finite system.
We then consider the thermodynamic limit as $N \to \infty$, where the state is normalizable and decaying exponentially as long as $|r_+|<1$
is satisfied. It is again remarked  that, in the limit as $N \to  \infty$, the effect of the modification at the right end  of the system should be negligible for the 
edge state localized at the left boundary.  
We therefore remove the tilde in the notations and arrive at the solution for the edge state at the left boundary 
$$
 \psi_{E=m_R}^{(m)} =  \mvec{\psi_{+,A}^{m}}{\psi_{+,B}^{m}}
$$
with
\begin{eqnarray*}
 \psi_{+,A}^{m} [n]&=& (-1)^{n-1} C_+ \cosh(q/2) r_+^{n-1},\\
 \psi_{+,B}^{m} [n] &=& (-1)^{n-1} C_+ e^{i\theta}\sinh(q/2) r_+^{n-1},
\end{eqnarray*}
where $\psi_{+,A(B)}^{m} [n]$ denotes the element of  $\psi_{+,A(B)}^{m}$ at the unit cell specified by the 
index $n$ and 
$$ 
 C_+=\bigg(\frac{1-|r_+|^2}{\cosh q}\bigg)^{1/2}
$$
is a normalization constant. 
This normalizable eigenstate reduces to the conventional edge state at the left end of the system in the limit as $q\to 0$ 
which resides only on the A sub-lattice.

The present approach to define the edge state also indicates that the state $\psi_{E=m_R}^{(m)}$ with the energy $m_R$ is inappropriate  
for the edge state at the right boundary, which should decay exponentially from right to left (${\rm Im}(k) <0$). 
Apparently, the condition (\ref{bc}) can not be removed by the local modification of the Hamiltonian at the left end 
of the system. We are therefore unable to construct a non-zero solution decaying from the right boundary to the left by 
assuming the form $ \psi_{E=m_R}^{(m)}$, suggesting that the state with the energy $m_R$ does not exist in the regime 
${\rm Im}(k) <0$.

To consider the edge states at the right boundary decaying exponentially toward the bulk, we examine   
the state $\psi_{E=-m_R}^{(m)}$ having the energy $-m_R$. With our choice of the unit cell, we have a B site 
at the right boundary. In such a case,
the condition (\ref{rightedge}) for the eigenstate $\psi_{E=-m_R}^{(m)}$ can be written,
using  the elements of $\phi_-^m={}^t (\phi_-^m[1],\ldots,\phi_-^m[N])$, as
$$
  (t+m_Re^{-i\theta}\sinh q) \phi_-^m[n] + t' \phi_-^m[n-1] = 0,
$$
for $n = 2, \ldots, N$ with 
\begin{equation}
 (t+m_Re^{-i\theta}\sinh q) \phi_-^m[1] =0
\label{bc2}
\end{equation}
at the left boundary. 
A non-zero solution is again prohibited by the condition (\ref{bc2})
for a finite system ($N < \infty$).

We then consider again a similar local modification of the Hamiltonian at the opposite (left) end of the system so that 
the coefficient of $\phi_-^m[1]$ in the condition (\ref{bc2}) becomes zero.
It is achieved by modifying the local Hamiltonian in the first cell ($n=1$) at the left end of the system as 
$$
 \tilde{H}^-_{\rm SSH} = H_{\rm SSH} -\{(m_Re^{-i\theta}\sinh q + t )a_1^\dagger b_1 + {\rm h.c.}\}.
$$
If we write the eigenstate $\tilde{\psi}_{E=-m_R}^{(m)}$ of the modified Hamiltonian $\tilde{H}^-_{\rm SSH}$ with the energy $-m_R$ as 
$$
 \tilde{\psi}_{E=-m_R}^{(m)} = T(q) \mvec{0}{\tilde{\phi}_-^m},
$$
the equations for  $\tilde{\phi}_-^m$  becomes 
$$
  (t+m_Re^{-i\theta}\sinh q) \tilde{\phi}_-^m[n] + t' \tilde{\phi}_-^m[n-1] = 0,
$$
for $n=2,\ldots,N$, which are exactly the same as those for $\phi_-^m$ except the condition (\ref{bc2}). 

We are then able to construct a solution 
that decays exponentially from the right edge to the bulk as
$\phi_-^m [n] \propto (-1)^{N-n} r _-^{N-n}$ with 
$$
 r_- \equiv \frac{t+m_Re^{-i\theta}\sinh q}{t'},
$$
which is normalizable in the limit  as $N\to \infty$ provided that $|r_-|<1$.
The edge state at the right boundary  is thus given by
$$
 \psi_{E=-m_R}^{(m)} =  \mvec{\psi_{-,A}^{m}}{\psi_{-,B}^{m}}
$$
with
\begin{eqnarray*}
 \psi_{-,A}^{m}[n]&=& (-1)^{N-n} C_- \cosh(q/2) r_-^{N-n},\\
 \psi_{-,B}^{m}[n] &=& (-1)^{N-n} C_- e^{i\theta}\sinh(q/2) r_-^{N-n}.
\end{eqnarray*}
Here $\psi_{-,A(B)}^{m} [n]$  denotes the element of $\psi_{-,A(B)}^{m}$ at the $n$-th unit cell and 
$$ 
 C_-=\bigg(\frac{1-|r_-|^2}{\cosh q}\bigg)^{1/2}
$$
is again a normalization constant. This solution reduces to the conventional edge state at the right boundary in the 
limit as $q \to 0$ which resides only on the B sub-lattice.

It is also noted that the state $ \psi_{E=-m_R}^{(m)}$ with the energy $-m_R$ can not describe the edge state 
at the left boundary decaying from left to right (${\rm Im}(k) >0$), because we can not remove the condition (\ref{bc2}) by a 
local modification of the Hamiltonian at the right boundary.  

The eigenstates with energies $E=\pm m_R$ are therefore indeed the edge states exponentially localized at the 
boundaries of the deformed system. These edge states exist only when the mass is smaller than 
the critical mass $m_c^\pm$ determined by $|r_\pm |=1$ so that 
condition $|r_\pm| <1$ is satisfied for $m<m_c^\pm$.

It is to be noted that $|r_+|$ and $|r_-|$ can be different, which means that the robustness of the 
edge state with the energy $E=m_R$ and that with $E=-m_R$ can be different for a deformed system.
As we shall see in the following, the edge state merges into the bulk band at a point where $|r_+|(|r_-|)$ becomes unity.  

Here we explicitly construct the edge state based on the boundary condition and its normalizability in a semi-infinite system. 
We note that  the edge states  with generic boundary conditions have been discussed 
for continuum models \cite{HKW, TDV,CKEH}. 

\subsection{Bulk versus Edge states}

As shown above, the edge state exhibits an instability at $|r_\pm|=1$. This condition can be understood as 
the point where the wave number of the plane wave solution with the energy $\pm m_R$ becomes real, which means that 
the edge state becomes one of the bulk states exactly at this point. 
This is the reason why the edge state merges into the bulk band at $|r_\pm| =1$. 
 
If we adopt the form $\phi_\pm^m[n] = e^{ikn}u(k)$, the conditions (\ref{leftedge}) and (\ref{rightedge}) for the states $\psi_{E=\pm m_R}^{(m)}$
are expressed as 
$$
 (t+t'e^{\pm ik}) \mp m_R e^{\pm i\theta} \sinh q = 0.
$$    
For $q=0$, this reduces to $t+t'e^{\pm ik}=0$ and the wave number $k$ is 
$k = \mp i \ln (-t/t')$. In the topological phase $(t'>t)$, the imaginary part of $k$ is positive (${\rm Im}(k) >0$) for the state  $\psi_{E=m_R}^{(m)}$
which corresponds to an exponentially decaying state at the left boundary. For the state  $\psi_{E=-m_R}^{(m)}$, we have ${\rm Im}(k) <0$ 
which means that the state is exponentially decaying from right to left at the right boundary. 
Note that the wave number is always complex independent of $m$, which 
means that the edge state never merges into the bulk band.

For $q\neq 0$, on the other hand,  the wave number $k$ of the states with $E=\pm m_R$ is given as
$$
 e^{\pm ik} = - \frac{t \mp m_Re^{\pm i\theta}\sinh q}{t'},
$$
and hence we have $e^{\pm ik} = -r_\pm$.  The requirement that $|r_\pm| =1$ is therefore equivalent to the condition 
that the wave vector $k$ is real.   At this point, the sign of the imaginary part of the wave number $k$ 
changes from positive(negative) to negative(positive) for the state $\psi_{E=m_R}^{(m)}(\psi_{E=-m_R}^{(m)})$, as the mass is increased.  
Since the state $\psi_{E=m_R}^{(m)}(\psi_{E=-m_R}^{(m)})$ exists only as the left(right) edge state, it is normalizable and therefore exists only when 
the imaginary part of the wave number ${\rm Im}(k)$ is positive(negative). 
In the deformed system, therefore, the edge states disappear at the critical mass $m_c^\pm$ where the imaginary part of $k$ vanishes, 
and becomes a bulk state there,  which does not happen in the conventional chiral symmetric systems.

\subsection{Numerical Results}

\begin{figure}[h]
\includegraphics[scale=0.33]{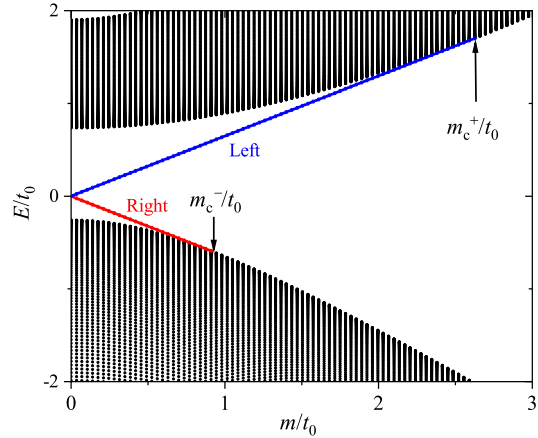}
\caption{(Color Online)
Energy eigenvalues for a system in the case of $\theta=0$ with the symmetry breaking term $m\Gamma $. The numerical results for a finite system with 200 unit cells having two open boundaries at each end of the system are shown. The 
parameters are assumed to be $t/t_0 = 0.65$, $t'/t_0=1.35$, and $q=1.0$. The bulk states in the Bloch bands are plotted by black symbols while 
the two edge states with the energy $\pm m_R$ in the bulk gap are plotted by red ($-m_R$) and blue ($+m_R$) lines. It is clearly seen 
that the edge state at the left end (blue line) exists in the range $0<m<(t'+t)/\tanh q$ and merges into the bulk band at $m_c^+ = (t'+t)/\tanh q \approx 2.6t_0$. 
The edge state at the right end (red line), on the other hand, merges into the bulk band at $m_c^-=(t'-t)/\tanh q \approx 0.9t_0$.
\label{energy_mass_q1}
}
\end{figure}

To confirm the analytical results, we perform numerical calculations in systems with boundaries. 
First, we consider the case  of  $\theta =0$, where the time-reversal symmetry (TRS) is preserved \cite{KAH}. 
In this case, the edge state with $E=m_R$ exists as long as
$$
  |t-m\tanh q| < |t'|,
$$
while the edge state with $E=-m_R$ exists when
$$
 |t+m\tanh q| < |t'|.
$$
For the case where $t$, $t'(>t)$, and $m$ are positive, these conditions leads to the fact 
that for $0<m\tanh q<t'-t$ the edge states at both  boundaries (left and right) exist in the gap while
for $ t'-t<m\tanh q<t+t'$ only the edge state at the left boundary can exist in the gap.  The critical masses $m_c^\pm$ 
over which the edge state merges into the bulk band are given by $m_c^+ = (t+t')/\tanh q$ and $m_c^- = (t'-t)/\tanh q$.
This can be clearly seen in Fig. \ref{energy_mass_q1} where the energy spectra for the deformed system 
$H_q^{\hat{x}} + m\Gamma$  with  open boundaries are shown.

\begin{figure}[h]
\includegraphics[scale=0.33]{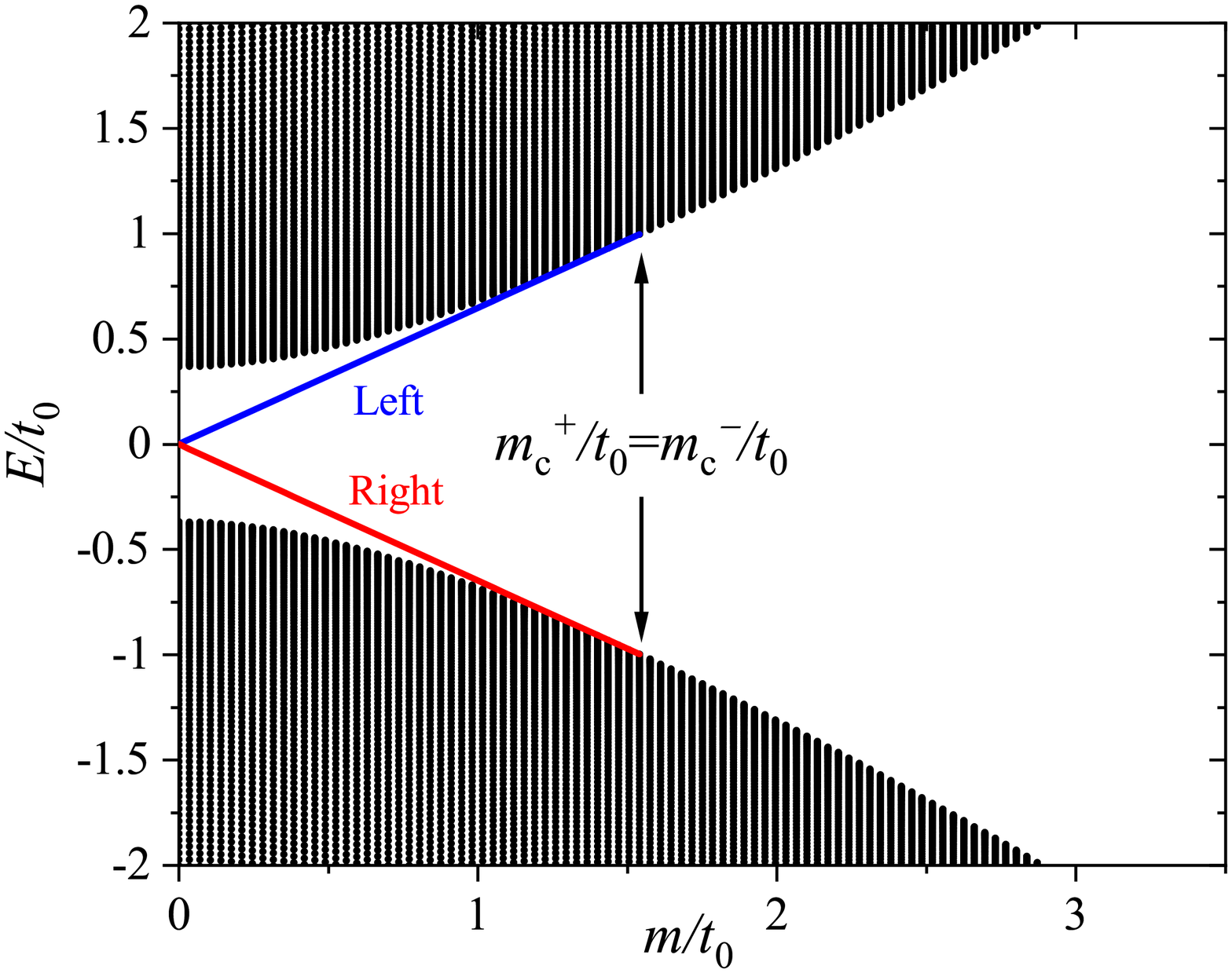}
\caption{(Color Online)
Energy eigenvalues for a system in the case of  $\theta =\pi/2$ with the symmetry breaking term $m\Gamma $. The numerical results for a finite system with 200 unit cells having two open boundaries at each end of the system is shown. The 
parameters are assumed to be $t/t_0 = 0.65$, $t'/t_0=1.35$, and $q=1.0$. The bulk states in the Bloch bands are plotted by black symbols while 
the two edge states with the energy $\pm m_R$ are plotted by red ($-m_R$) and blue ($+m_R$) lines. It is clearly seen 
that both the edge states (blue and red lines ) exist in the range $0<m<\sqrt{t'^2-t^2}/\tanh q$ and merges into the bulk band at $m_c^+ =m_c^- =  \sqrt{t'^2-t^2}/\tanh q \approx 1.55 t_0$. 
\label{energy_mass_q1_y}
}
\end{figure}

Next, we consider the case of $\theta =\pi/2$, where the time-reversal symmetry is broken.
The critical mass $m_c^\pm$ is then given by 
$m_c^+=m_c^- = \sqrt{t'^2-t^2}/\tanh q$, and therefore both edge states exist only when the condition $m < m_c^\pm$ is satisfied (Fig. \ref{energy_mass_q1_y} ).
In this case, the energy spectra are symmetric with respect to $E=0$. 
Note that in both cases, no eigenstate with the energy $E=\pm m_R$ exists for $m>m_c^\pm$.   

\section{Summary}
We have shown that the bulk-edge correspondence in topological phases can be extended to 
the systems without the conventional chiral symmetry but respecting the generalized chiral symmetry.
Systems respecting the generalized chiral symmetry  are generated by the algebraic deformation which preserves the 
bulk topological invariants of the original chiral symmetric system as well as the zero-energy edge states at the boundaries. 
We have explored the bulk-edge correspondence in a deformed Su-Schrieffer-Heeger model in one dimension. We have found  
interestingly that the edge states in 
a deformed system with generalized chiral symmetry exhibit an instability when the symmetry is broken by the mass.  
The edge state disappears at a certain critical value of the mass, where it touches to the bulk band and 
becomes a bulk state with a real wave number, which never happens for the conventional chiral symmetric systems. 
The present analysis suggests that in the bulk-edge correspondence with the generalized chiral symmetry, 
the edge states  are adiabatically  connected to the bulk states by modifying
the strength of the symmetry breaking.

\begin{acknowledgments}

The work was supported in part by JSPS KAKENHI grant numbers 
JP19K03660 (TK), 
and JP17H06138. 

\end{acknowledgments}
\appendix

\section{Deformation and Lorentz boost}
The present deformation has the same hyperbolic form as the Lorentz boost \cite{LSB,TCG,TJIBPCG,TJIBPCG2}.
To see the relationship between them, we consider the two-dimensional massless Dirac electrons
described by the effective Hamiltonian $H = v_F(\sigma_x p_x+\sigma_y p_y)$. Then the 
Schr\"{o}dinger equation becomes 
$$
 (i\hbar\frac{\partial }{\partial t} - H) \psi =0.
$$ 
This can be reduced to, in the real-space representation,
$$
 (\partial_0 +\sigma_x \partial_x +\sigma_y\partial_y ) \psi =0, 
$$
where $\partial_0 = \frac{\partial }{\partial x_0}$, $\partial_x = \frac{\partial }{\partial x}$, and $\partial_y = \frac{\partial }{\partial y}$ with 
$x_0 = v_F t$. We then consider a Lorentz boost for the frame moving in the $x$ directions with a velocity $v$. The coordinates 
$(x_0',x',y')$ in such a frame
are given by
$$
 \mvec{x_0'}{x'} = \mmat{\cosh q}{-\sinh q}{-\sinh q}{\cosh q} \mvec{x_0}{x}  ,\ y'=y,
$$
where the parameter $q$ is given by $\tanh q = \frac{v}{c}$ with the speed of light $c$.
In this frame, the Schr\"{o}dinger equation becomes
$$
 \exp(q\sigma_x/2) [\partial_0' +\sigma_x\partial_x'+\sigma_y \partial_y'] \exp(q\sigma_x/2) \psi =0,
$$
with which we arrive at  $[\partial_0' +\sigma_x\partial_x'+\sigma_y \partial_y'] \psi' =0$ with $\psi' =e^{q\sigma_x/2} \psi$, 
where $\partial_0' = \frac{\partial }{\partial x_0'}$, $\partial_x' = \frac{\partial }{\partial x'}$, and $\partial_y' = \frac{\partial }{\partial y'}$.
The present deformation for the operator $[\partial_0 +\sigma_x\partial_x+\sigma_y \partial_y] $ therefore corresponds to the Lorentz boost with
$\tanh q = v/c$.

\section{Winding number of Deformed systems}
Here we  consider the deformation 
$$
 H_\tau(q) = T_\tau(q)^{-1} H_{\rm SSH} T_\tau(q)^{-1},
$$
with $\bm{\tau} = (\cos \theta,\sin\theta,0)$,
which yields in the momentum space 
$$
 H_\tau(q) = -\sinh q \ {\rm Re}[e^{i\theta}d(k)] I_2 + \mmat{0}{d_{q,\theta} (k)}{d_{q,\theta}^*(k)}{0},
$$ 
where 
$d_{q,\theta}(k) = e^{-i\theta}\{{\rm Re}[e^{i\theta}d(k)] \cosh q +i {\rm Im}[e^{i\theta}d(k)]\}$ with $d(k) = t+t'e^{-ik}$ and  
$I_2$ stands for the 2 $\times$ 2 identity matrix.
The energy eigenvalues are given  by 
$$
 E_{q,\pm}^\tau = 
 -\sinh q\  {\rm Re}[e^{i\theta}d(k)]\pm \sqrt{|d_{q,\theta}(k)|^2},
$$
and 
the corresponding eigenstates $|\psi_\pm^\tau \rangle$ with $H_\tau(q) |\psi_\pm^\tau \rangle$ $= E_{q,\pm}^{\tau}
 |\psi_\pm^\tau \rangle$ are  given by
$$
 |\psi_\pm^\tau \rangle = \frac{1}{\sqrt{2}}\mvec{\pm \alpha_\tau^*}{1}, \quad \alpha_\tau^* = \frac{d_{q,\theta}(k)}{|d_{q,\theta}(k)|} .
$$
The $Q$-matrix \cite{CTSR} is therefore given by
$$
 Q = |\psi_+^\tau\rangle \langle \psi_+^\tau | - | \psi_-^\tau\rangle \langle \psi_-^\tau | 
 = \mmat{0}{\alpha_\tau^*}{\alpha_\tau}{0}.
$$
The winding number $w$ can then be defined by 
$$
 w = \frac{i}{2\pi}\int_{\rm BZ} d\alpha_\tau \ \alpha_\tau^{-1}.
$$
This winding number becomes non-zero when the trajectory of $d_{q,\theta}(k)$ in the complex plane 
encircles the origin when $k$ moves from $0$ to $2\pi$.  For $q=0$, $d_{q,\theta }(k)$ becomes $d(k)$ and it encircles the 
origin when $t'>t$ \cite{CTSR,RH}.
In the present deformed systems with $q\neq 0$, $d_{q,\theta}(k)$ is simply scaled by a factor $\cosh q$ in the direction determined by 
$\theta$ and hence 
the winding number is the same as $d(k)$. The winding number is therefore an invariant of the present deformation, with which we conclude that the topological phase in the deformed systems is always given by $t'>t$ and the bulk-edge correspondence is valid independent of $q$.

\section{Deformed SSH Models with/without Time-Reversal Symmetry}

\begin{figure}[h]
\includegraphics[scale=0.8]{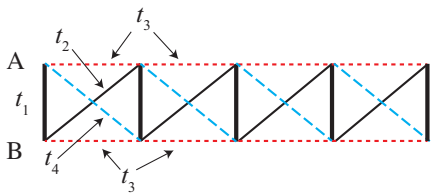}
\caption{(Color Online)
The deformed SSH  model with $\bm{\tau}=\hat{\bm{x}}$ in real space. 
Thick solid lines (black), thin solid lines (black), dotted lines (red), and dashed lines (blue) represent 
hopping amplitudes $t_1$, $t_2$, $t_3$, and $t_4$, respectively.    
\label{ssh_deform_x}
}
\end{figure}

Here we show typical cases of deformed SSH models with/without the time-reversal symmetry.
When we assume $\bm{\tau} = \hat{\bm{x}} = (1,0,0)$, the deformed SSH model 
$$
 H_{\hat{x}}(q) = \exp(-q\sigma_x /2) H_{\rm SSH}  \exp(-q\sigma_x /2)
$$
respects the 
time-reversal symmetry. In this case, 
a next nearest-neighbor and a 3rd nearest-neighbor transfer integrals  as well as 
a uniform energy shift emerge in the deformed Hamiltonian in the real space,  which   is 
equivalent to a ladder Hamiltonian described by 
\begin{eqnarray*}
 H_q^{\hat{x}} &= &\sum_n \varepsilon (a_n^\dagger a_n+b_n^\dagger b_n)
 + \sum_n t_1\  a^\dagger_n b_n 
 +t_2 \  a_{n+1}^\dagger b_n \\
 &+ &t_3 (a_{n+1}^\dagger a_n +b^\dagger_{n+1}b_n) 
 + t_4 \ a^\dagger_{n-1}b_n + {\rm h.c.}
\end{eqnarray*}
where $\varepsilon=-t\sinh q$, $t_1 = t\cosh q$, $t_2 = t'(\cosh q +1)/2$, $t_3 = - t'(\sinh q)/2$, $t_4 = t'(\cosh q -1)/2$  (Fig. \ref{ssh_deform_x}).

\begin{figure}[h]
\includegraphics[scale=0.6]{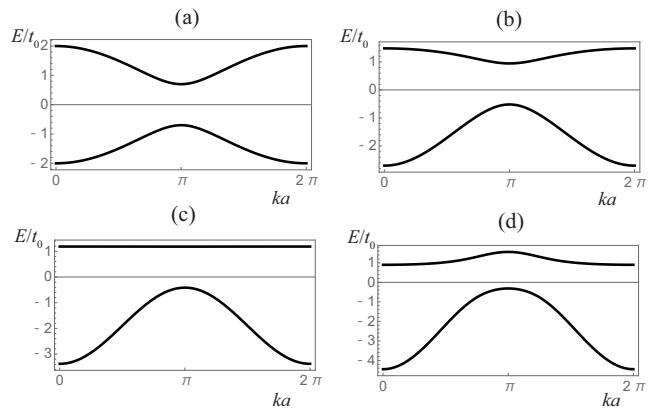}
\caption{
Energy dispersions $E_{q,\pm}^{\hat{x}}$ of the deformed SSH model for (a) $q=0$, (b) $q=0.3$, (c) $q=\tanh^{-1}(t/t')\approx 0.524$, and (d) $q=0.8$. Because of the time-reversal symmetry, we have $E_{q,\pm}^{\hat{x}} (k) = E_{q,\pm}^{\hat{x}}(-k)$. 
Here we assume $t = t_0-\delta t$, $t'=t_0+\delta t$ with $\delta t/t_0 = 0.35$ so that the original SSH model is topologically non-trivial.
The lattice constant of the corresponding lattice is 
denoted by $a$. A flat band appears for the case of (c). For $q>\tanh^{-1}(t/t')$, the band gap becomes indirect (d).
\label{ssh_deform_x_disp}
}
\end{figure}

In this series of deformed topological SSH models, we find at $q=\tanh^{-1}(t/t')$ that the deformed model has a flat band. Note that this flat band model appears 
only when the original SSH model is topologically non-trivial ($t'>t$). In fact, for the case  $\tanh q = t/t'$, we have 
$$
 E_{q,+}^{\hat{x}}  =  \sqrt{t'^2-t^2}, \quad E_{q,-}^{\hat{x}}= \frac{-(t'^2+t^2)-2tt'\cos k}{ \sqrt{t'^2-t^2}},
$$ 
and hence the energy of the flat band is $\sqrt{t'^2-t^2}$. In Fig. \ref{ssh_deform_x_disp}, we show examples of the energy dispersions 
of the deformed SSH model, where the energy bands become asymmetric for 
$q\neq 0$. 
Models generated by the present  deformation are however topologically non-trivial for any $q$.
The energy gap, which can be indirect (Fig. \ref{ssh_deform_x_disp} (d)), never closes in the deformation.

\begin{figure}[h]
\includegraphics[scale=0.25]{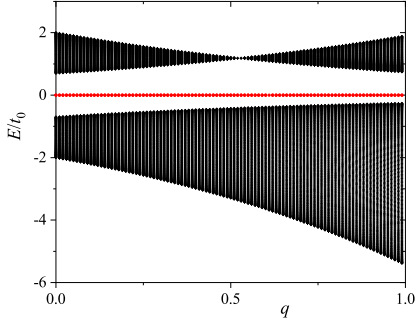}
\caption{(Color Online)
Energy eigenvalues for a finite system with 200 unit cells having two open boundaries at each end of the system. The 
parameters are assumed to be $t/t_0 = 0.65$ and  $t'/t_0=1.35$. The bulk states in the Bloch bands are plotted by black symbols while 
the two edge states at $E=0$, located either the left or the right ends of the system, are plotted by red symbols.  
\label{energy_open}
}
\end{figure}

The bulk-edge correspondence in the present models can be confirmed by the appearance of the edge states when the 
system has open boundaries. We show in Fig. \ref{energy_open} the energy eigenvalues of a finite system with 
open boundaries at each end of the system. 
It is clearly seen that the edge states exist  exactly at zero energy throughout the deformation.

\begin{figure}[h]
\includegraphics[scale=0.8]{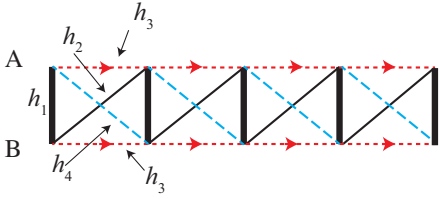}
\caption{(Color Online)
The SSH  model deformed by $\sigma_y$ in real space. Thick solid lines (black), thin solid lines (black), dotted lines (red), and dashed lines (blue) represent 
hopping amplitudes $h_1$, $h_2$, $h_3$, and $h_4$, respectively.  The hopping $h_3$ becomes imaginary  while the others 
are real. 
\label{ssh_deform_y}
}
\end{figure}

Next, we consider the models with $\bm{\tau} = \hat{\bm{y}} =  (0,1,0)$ as 
$$
 H_{\hat{y}} (q)= \exp(-q\sigma_y/2) H_{\rm SSH} \exp(-q\sigma_y/2),
$$
where the time-reversal invariance is broken for $q\neq 0$.
In the real space, the Hamiltonian can be written again in a ladder system as (Fig.\ref{ssh_deform_y})
\begin{eqnarray*}
 H^{\hat{y}}_q &=& \sum_n h_1a_n^\dagger b_n + h_2 a_{n+1}^\dagger b_n \\
 & +& h_3(a_{n+1}^\dagger a_n + b_{n+1}^\dagger b_n) +h_4a_{n-1}^\dagger b_n + {\rm h.c.},
\end{eqnarray*}
where $h_1 = t$, $h_2=t'(\cosh q+1)/2$, $h_3 = -it'(\sinh q)/2$, $h_4=-t'(\cosh q-1)/2$.
Because of the breaking of the time-reversal invariance, the hopping $t'_3$ becomes 
imaginary.

\begin{figure}[h]
\includegraphics[scale=0.6]{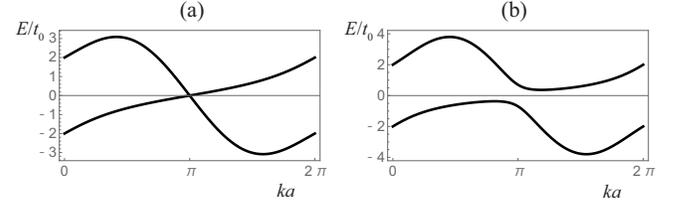}
\caption{
Examples of the energy dispersion $E^{\hat{y}}_{q,\pm}(k)$ for $q=1.0$. The critical case (a) $t/t'=1.0$ and the topological case (b) 
$t/t_0 = 0.65$ and $t'/t_0 = 1.35$ are presented. Although the time-reversal symmetry is broken,  we have a relation $E^{\hat{y}}_{q,\pm}(k) = E_{-q,\pm}^{\hat{y}}(-k)=E_{q,\mp}^{\hat{y}}(-k)$.
\label{E_k_plot2}
}
\end{figure}

Examples of the dispersion relations for the deformed systems are shown in Fig. \ref{E_k_plot2}, where $E_{q,\pm}^{\hat{y}}(k) \neq E_{q,\pm}^{\hat{y}}(-k)$
due to the breaking of the time-reversal invariance.

\begin{figure}[h]
\includegraphics[scale=0.25]{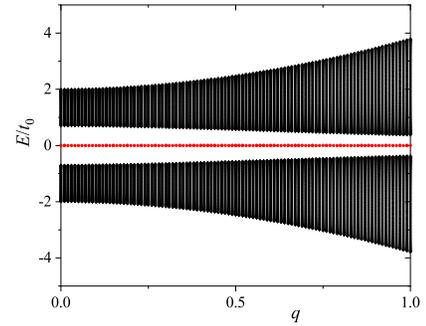}
\caption{(Color Online)
Energy eigenvalues for a finite system with 200 unit cells having two open boundaries at each end of the system. The 
parameters are assumed to be $t/t_0 = 0.65$ and  $t'/t_0=1.35$. The bulk states in the Bloch bands are plotted by black symbols while 
the two edge states at $E=0$ are plotted by red symbols.  
\label{energy_open_y}
}
\end{figure}

The appearance of the edge states at  $E=0$ for an open system is also confirmed for $t'>t$.
In Fig.\ref{energy_open_y}, we show the energy eigenvalues of an open system with 200 unit cells, where
the spectra are symmetric with respect to $E=0$. The edge states at $E=0$ again exist for any value of $q$.

\begin{thebibliography}{10}
\bibitem{HK} M.Z. Hasan and C.L. Kane, Rev. Mod. Phys. {\bf 82}, 3045 (2010).
\bibitem{QZ} X.-L. Qi and S.-C. Zhang, Rev. Mod. Phys. {\bf 83}, 1057 (2011).
\bibitem{Hatsugai} Y. Hatsugai, Phys. Rev. Lett. {\bf 71}, 3697 (1993).
\bibitem{AZ} A. Altland and M.R. Zirnbauer, Phys. Rev. B {\bf  55}, 1142 (1997).
\bibitem{RSFL} S. Ryu, A. P. Schnyder, A. Furusaki and A.W.W. Ludwig, New J. Phys. {\bf 12}, 065010 (2010).
\bibitem{HA1} Y. Hatsugai and H. Aoki, in {\it Physics of Graphene}, edited by H. Aoki and M.S. Dresselhaus (Springer, Heidelberg and New York, 2014), P. 213.
\bibitem{CTSR} C-K. Chiu, J.C.Y. Teo, A.P. Schnyder, and S. Ryu, Rev. Mod. Phys. {\bf 88}, 035005 (2016).
\bibitem{RH} S. Ryu and Y. Hatsugai, Phys. Rev. Lett. {\bf 89},077002 (2002).

\bibitem{YW} S. Yao and Z. Wang, Phys. Rev. Lett. {\bf 121}, 086803 (2018).
\bibitem{YM} K. Yokomizo and S. Murakami, Phys. Rev. Lett. {\bf 123}, 066404 (2019). 
\bibitem{KHMA} T. Kawarabayashi, Y. Hatsugai, T. Morimoto, and H. Aoki, Phys. Rev. B{\bf 83}, 153414 (2011); Int. J. Mod. Phys.: Conf. Series {\bf 11}, 145 (2012).
\bibitem{HKA} Y. Hatsugai, T. Kawarabayashi, and H. Aoki, Phys. Rev. B{\bf 91}, 085112 (2015).
\bibitem{KAH} T. Kawarabayashi, H. Aoki, and Y. Hatsugai, Phys. Rev. B{\bf 94}, 235307 (2016).
\bibitem{TSTNK} N. Tajima, S. Sugawara, M. Tamura, Y. Nishio, and K. Kajita, J. Phys. Soc. Jpn. {\bf 75}, 051010 (2006).
\bibitem{KKS} S. Katayama, A. Kobayashi, and Y. Suzumura, J. Phys. Soc. Jpn. {\bf 75}, 054705 (2006); {\bf 75}, 023708 (2006).
\bibitem{KKSF} A. Kobayashi, S. Katayama, Y.Suzumura, and H. Fukuyama, J. Phys. Soc. Jpn. {\bf 76}, 034711 (2007).
\bibitem{KSFG} A. Kobayashi, Y. Suzumura, H. Fukuyama, and M.O. Goerbig, J. Phys.Soc. Jpn. {\bf 78}, 114711 (2009).
\bibitem{KNTSK} K. Kajita, Y. Nishio, N. Tajima, Y. Suzumura, and A. Kobayashi, J. Phys. Soc. Jpn. {\bf 83}, 072002 (2014).
\bibitem{GFMP} M.O. Goerbig, J.-N. Fuchs, G. Montambaux, and F. Pi\'{e}chon, Phys. Rev. B {\bf 78}, 045415 (2008).
\bibitem{MHT} T. Morinari, T. Himura and T. Tohyama, J. Phys. Soc. Jpn. {\bf 78}, 023704 (2009).
\bibitem{NN} H.B. Nielsen and M. Ninomiya, Nucl. Phys. B {\bf185}, 20 (1981).
\bibitem{HFA} Y. Hatsugai, T. Fukui, and H. Aoki, Phys. Rev. B {\bf 74}, 205414 (2006);
 Eur. Phys. J. Special topics, {\bf 148}, 133 (2007).
\bibitem{Hatsugai2} Y. Hatsugai, J. Phys. Conf. Series {\bf 334}, 012004 (2011).
\bibitem{KAH2} T. Kawarabayashi, H. Aoki, and Y. Hatsugai, Phys. Status Solidi B {\bf256}, 1800524 (2019).

\bibitem{SSH} W.P. Su, J.R. Schrieffer, and A.J. Heeger, Phys. Rev. Lett. {\bf 42}, 1628 (1979); Phys. Rev. B{\bf 22}, 2099 (1980).
\bibitem{JR} R. Jackiw and C. Rebbi, Phys. Rev. D {\bf 13}, 3398 (1976).  
\bibitem{SS} W.P. Su and J.R. Schrieffer, Phys. Rev. Lett. {\bf 46}, 738 (1981). 
\bibitem{GW} J. Goldstone and F. Wilczek, Phys. Rev. Lett. {bf 47}, 986 (1981).
\bibitem{RM} M.J. Rice and E.J. Mele, Phys. Rev. Lett. {\bf 49}, 1455 (1982).
\bibitem{JS} R. Jackiw and G. Semenoff, Phys. Rev. Lett. {\bf 50}, 439 (1983).
\bibitem{HKSS} A.J. Heeger, S. Kivelson, J.R. Schrieffer, and W.P. Su, Rev. Mod. Phys. {\bf 60}, 781 (1988).
	  
\bibitem{MAG} E.J. Meier, F.A. An, and B. Gadway, Nature Commun. {\bf 7}, 13986 (2016); Phys. Rev A{\bf 93}, 051602(R) (2016).
\bibitem{DOHL} R. Drost, T. Ojanen, A. Harju, and P. Liljeroth, Nat. Phys. {\bf 13}, 668 (2017).
\bibitem{HKODL} M.N. Huda, S. Kezilebieke, T. Ojanen, R. Drost, and P. Liljeroth, Quantum Materials {\bf 5}, 17 (2020).

\bibitem{HKW} K. Hashimoto, T. Kimura, and X. Wu, Prog. Theor. Exp. Phys. {\bf 2017}, 053I01 (2017); Prog.  Theor. Exp. Phys. {\bf 2019}, 029201 (2019).
\bibitem{TDV} C. Tauber, P. Delplace and A. Venaille, Phys. Rev. Research {\bf 2}, 013147 (2020).
\bibitem{CKEH} D.R. Candido, M. Kharitonov, J.C. Egues, and E.M. Hankiewicz, Phys. Rev. B{\bf 98}, 161111(R) (2018). 

\bibitem{LSB} V. Lukose, R. Shankar, and G. Baskaran, Phys. Rev. Lett. {\bf 98}, 116802 (2007).
\bibitem{TCG} S. Tchoumakov, M. Civelli, and M.O. Goerbig, Phys. Rev. B{\bf 95}, 125306 (2017).
\bibitem{TJIBPCG} S. Tchoumakov, V. Jouffrey, A. Inhofer, E. Bocquillon, B. Pla\c{c}ais, D. Carpentier, and M.O. Goerbig, Phys. Rev. B{\bf 96}, 201302(R) (2017).
\bibitem{TJIBPCG2} A. Inhofer, S. Tchoumakov, B.A. Assaf, G. F\`{e}ve, J. M. Berroir, V. Jouffrey, D. Carpentier, M.O. Goerbig, B. Pla\c{c}ais, K. Bendias, D.M. Mahler, E. Bocquillon, R. Schlereth, C. Br\"{u}ne, H. Buhmann, and L.W. Molenkamp, Phys. Rev. B{\bf 96}, 195104 (2017).

\end{thebibliography}

\vfill
\end{document}